# Applying MVC and PAC patterns in mobile applications

Plakalović D., Simić D.

**Abstract-**Additional requirements are set for mobile applications in relation to applications for desktop computers. These requirements primarily concern the support to different platforms on which such applications are performed, as well as the requirement for providing more modalities of input/output interaction. These requirements have influence on the user interface and therefore it is needed to consider the usability of MVC (Model-View-Controller) and PAC (Presentation-Abstraction-Control) design patterns for the separation of the user interface tasks from the business logic, specifically in mobile applications. One of the questions is making certain choices of design patterns for certain classes of mobile applications. When using these patterns the possibilities of user interface automatic transformation should be kept in mind. Although the MVC design pattern is widely used in mobile applications, it is not universal, especially in cases where there are requirements for heterogeneous multi-modal input-output interactions.

—————— ◆ ——————

## 1 INTRODUCTION

THE construction of interactive software systems with multiple types of user interfaces is expensive and subject to errors, when the user interface is closely interwoven with the functional core. In this case, a separate application is made for each type of user interface applications and the non-interface-specific code is duplicated in each application. This results in duplicate efforts in implementation, which according to [1] usually result in "copy & paste" variety. Of course, the increased complexity of implementation is transferred onto the increasing complexity of testing and maintenance, and it is also known that the update of the copies is inevitably imperfect. "Slowly, but surely, the applications expected to provide the same functional core are developing into various systems" [1]. The result is the expansion of future changes on many modules.

However, the above-mentioned problems can be avoided by the separation of the user interface tasks from the functional core. When solving this problem two software patterns may be used and they are related to the way of building the applications user interface. These are the "Model-View-Controller (MVC) and Presentation-Abstraction-Control (PAC). The next section of this paper will define the problem of choice between MVC and PAC software pattern when designing mobile applications. The third and fourth section will display information from the literature on these software patterns. The fifth, central section will establish the connection between the conditions facing mobile application and the selection of a particular software pattern for the user interface architecture.

## 2 DEFINITION OF THE PROBLEM

When defining requirements set before mobile applications, let us start from a vision of pervasive computing which allows users access to relevant applications and data on any location on any device, in a way that is adapted to the user and to the task that he is currently performing. Fundamental characteristics of the pervasive applications are mobility and context awareness. One consequence of mobility is that applications must work on various types of devices, including devices installed in different environments and devices that users carry with them [3]. The need for context-aware applications arises from the requirement to use the applications in contexts that are different compared to the computer workstation with keyboard, mouse and screen. The pervasive computing application users are typically focused on another task, and not on the very use of the mobile computer device, and may even be unaware that they use a computer device. Applications must adapt to interact with the user in a way that is appropriate for the current user context and activities, taking advantage of locally available equipment, without disturbing the user in performing his current task [4].

Let us consider an example of a pervasive calendar application that should have the following properties. First, the application will be able to execute on multiple platforms, from a networked phone (with limited user interface, but still connected) to the personal digital assistant, PDA (Personal Digital Assistant) with a richer user interface and larger bandwidth range, but not always linked, to the conference indoor computer (with a rich user interface and high bandwidth range and which is always linked). Furthermore, the user will be able to interact with the application by using multiple interface modalities, such as graphic user interface (GUI), voice interface, or a combination of these two. Second, the application will be sensitive to the environment in which it is performed: for example, at home, the application displays the family calendar as a default. If the application is used in the office and if the user is almost too late for the next meeting, the application can display the business calendar with information about his next meeting even with the emphasis on this information.

The user interface capabilities include output capabilities (e.g., size and colour); input capabilities, such as the number of buttons, rollers and other controls, as well as software accessories available to manipulate the input and output capabilities. It is necessary to re-write the component view for each device because of differences in these capabilities from one device to another. In some cases, the structure of Views will also influence the structure of the Controller.

## 3 SOFTWARE PATTERN "MODEL-VIEW-CONTROLLER" - MVC

According to the MVC software pattern, the application should have at least three components. The Model component includes the core of application data and logic domain functionality. The View obtains data from the Model and displays them to the user. The Controller receives and interprets input into the requirements for the Model or the View.

One of the first uses of the MVC software pattern in object-oriented software applications was in the SmallTalk programming language. The description of the tasks of specific components in this implementation and the manner of mutual co-ordination are given in [2].

The Model represents the data of the domain problem for which the application is intended and business rules that govern data access and update. Since the Model must be independent, it cannot have a reference either to the View or to any part of the application Controller.

The View component is responsible for presenting information to the user. Different Views represent information from the Model in different ways. Each View defines the procedure of update of information display to the user, which is activated by the change propagation mechanism. When such procedure is called, the View searches the current value of data that will show from the Model and puts them on the screen. During initialization, all Views join the Model and are registered by the change propagation mechanism. Each View creates an appropriate Controller and one Controller matches one View. The Views often provide functionality that allows Controllers to manipulate with the display. It is usable for the operations started by the user and which do not affect the Model, such as scrolling, for example.

Consistency of Views with the Model is provided in a way that the Model classes define the change notification mechanism, usually using the Observer pattern that is applied in Smalltalk as well. This allows the View and Controller components to be informed about changes in the Model. Since these components register themselves with the Model and the Model has no knowledge of any particular View or the Controller, it does not violate the independence of the Model. The notification mechanism is the one that enables immediate update, which is characteristic of MVC application with graphic user interface.

The View obtains data from the Model and displays them to the user. The view represents the exit from the application and it reflects the Model content. The View accedes to the application data through the Model and specifies the manner in which the data should be displayed. The Model responsibility is to maintain the consistency of its presentation with the Model changes. The Controller transfers interactions with the user into the actions to be carried by the Model. In the standalone GUI client, user interactions may be a click of a button or a menu choice, while in the Web application they appear as GET and POST HTTP requests. Actions performed by the Model include activating business processes or changes of the Model status. The Controller responds by choosing the appropriate View based on user interactions and exits from the actions performed by the Model.

The View generally has free access to the Model, but it must not change the Model status. The Views only read the shows of the Model status. The View reads the data from the Model by using the inquiry call method whose execution is provided by the Model.

The Controller component accepts user input as events. The manner of delivery of events towards the Controller depends on the user interface platform. For simplicity, let us assume that each Controller implements procedure for handling events that is called for any relevant event. Events are translated into requirements on the Model or the associated View. If the behaviour of the Controller depends on the Model status, the Controller will be registered with the change propagation mechanism and will implement the update procedure. For example, this is necessary when the Model change allows or prevents an item on the menu. The class diagram in Figure 1 and sequence diagram in Figure 2 show the application of MVC architectural petterns in the application implemented in the Java 2 Mobile Edition platform. The application consists of MIDlet represented by the Controller [5], View class and Model class. Upon application start, the Controller creates models and views instances. Views then register with the Model. The Controller class accepts user commands and calls the operation to set data in the Model. Then, using Observer pattern, the Model calls the update method () of all registered Views in order to reflect on the Views all changes that were made in the Model.

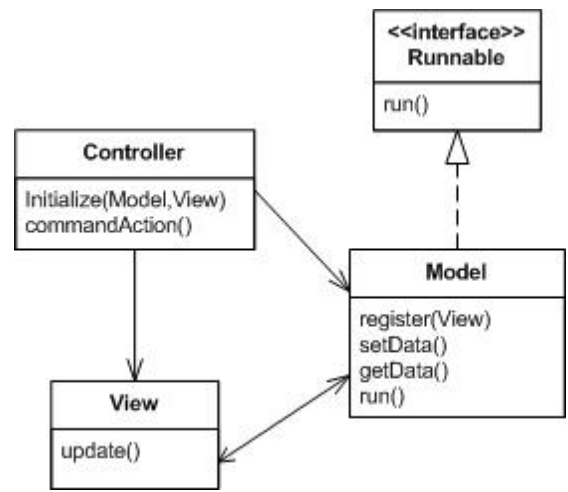

Figure1: Classes diagram as per MVC pattern in Java 2 Mobile Edition

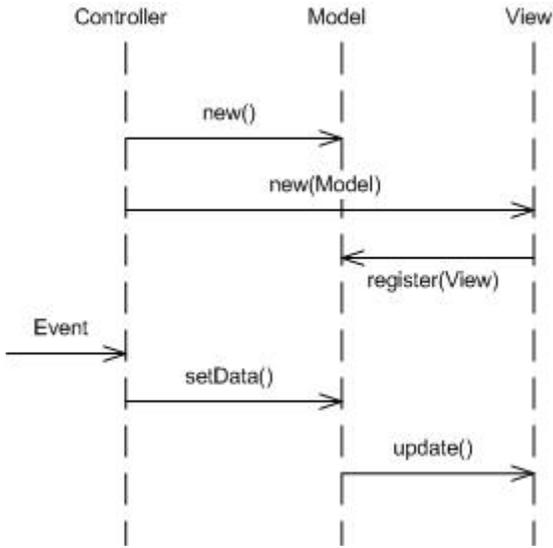

Fig. 2. Sequence diagram as per MVC pattern in Java 2 Mobile Edition Platform

In the MVC pattern, the Controller is not an intermediary between Views and Models, and it does not sit between Models and Views. The separation of the tasks of the Models and Views can be achieved by using the Observer pattern, not through the Controller [6].

The separation of control aspects from the views, allows a combination of several different controllers with one view. This flexibility can be used to implement different operation modes, such as, for example, "beginner user" versus "expert user", or for designing a read-only view by using the controller that ignores any input.

### 3.1 Benefits of MVC

The application of MVC patterns allows multiple views over a single model. MVC strictly separates the model from the user interface components and therefore multiple views can be implemented and used with a single model. During execution, multiple views can be open at the same time, and the views can also be opened and closed dynamically. Second, MVC allows the view synchronization. Model change propagation mechanism provides that all views registered with the object of "observer" are notified about changes of application data exactly on time. This allows synchronization of all dependent views and controllers.

MVC also allows creation of "Pluggable" views and controllers. Conceptual separation of MVC allows to finds a replacement for view and controller object. User interface objects can even be replaced during the time of execution. The fourth benefit relates to the possibility of "look and feel" change of the user interface characteristics. Since the model is independent of the user interface code, transferring of the MVC application to a new platform does not affect the functional application core. The appropriate implementations of the views and controllers components for each platform are only required.

### 3.2 MVC pattern disadvantages

Strict monitoring of MVC structure is not always the best way to build interactive application because the use of separate components of Models, Views and Controllers for the menus and simple text elements increases the complexity without getting a lot of flexibility. The use of this pattern also creates the potential for an excessive number of updates. If actions of one user result in many updates, the model needs to skip the unnecessary notification of changes. The case may be that all views are not interested in the propagation of all changes in the model. For example, the view with the window and icons might not need to be updated as long as its window is not restored to its normal size. The lack of closeness between views and controllers is also a disadvantage because it prevents their specific use. It is not likely that the view will be used without its controller, or vice versa, with the exception of "read-only" views that share a common controller that ignores all input in the event of operation with "read-only" view. Further, there is a close connection between the view and the controller with the model. Both view and controller components make direct calls to the model. This implies that changes in the interface model probably break the code both with the view and the controller. This problem is increased if the system uses a large number of views and controllers. The inevitability of changes in the View and Controller components also appears as an advantage, at their transfer to another platform. All dependence of the user interface platform is encapsulated within the View and Controller components. However, both components also contain the code that is independent of specific platforms. The MVC system transfer requires the separation of platform dependent code before re-writing, in order to encapsulate platform dependencies.

## 4 SOFTWARE PATTERN "PRESENTATION-ABSTRACTION-CONTROL"-PAC

PAC design pattern is based on the concept of cooperative agents, which are organized in a hierarchical structure. Each agent is a unitary aspect of the system, which works as a node in the hierarchy of agents and consists of the components of Presentation, Abstraction and Control.

According to this pattern, interactive systems consist of cooperative agents. One type of agents specialized in human-computer interaction accepts user input and displays data. Other agents maintain the system data model and provide functionality that is based on such data. Additional agents are responsible for different tasks such as error management or communication with other software systems.

This design pattern has emerged under the influence of the following circumstances [2]:
- Agents often have a need to maintain their own status and data. In order to ensure the execution of the entire

application task, agents must cooperate effectively, which requires a data, message and event exchange mechanism.
- Interactive agents provide their own user interface, since their human-computer interaction often differs a lot.
- Systems develop over time. Their presentation aspect is particularly subject to change. In addition, changes to individual agents or expansion of the system with new agents, should not affect the entire system.

The solution is in the structuring of interactive applications as in the form of a tree whose nodes are PAC agents. There should be one top agent, a few middle-level agents and a few more agents at the bottom of the hierarchy. Each agent is responsible for certain aspect of the functionality of applications and consists of three components: presentation, abstraction and control. The entire hierarchy reflects transitive dependencies among agents. Each agent depends on the higher-level agents in the hierarchy, up to the top agent.

The agent's Presentation component provides a visible behaviour of the PAC agent. The Abstraction component maintains data model that is below the agent and it provides functionality that works with such data. The Control Component links components of Presentation and Abstraction and provides functionality that allows the agent to communicate with other PAC agents [7].

The top PAC agent provides the functional core of the system. Most of other PAC agents depend or operate with such core. The lowest-level PAC agents are semantic concepts with which system users can operate. In addition, they support all operations that users can carry out with such concepts. Middle-level PAC agents are either combinations or connections among lower-level agents. For example, the middle-level agent can maintain several views with the same data.

### 4.1 Structure

The main responsibility of the top level PAC agent is to provide a global data model for software. This task is performed by the Abstraction component in the top agent. The Abstraction component interface provides functions for manipulation with the data model and for information search. The representation of data within the Abstraction component must be independent of the medium that supports the adaptation of the PAC agent for different environments without major changes in its Abstraction component.

The Presentation component of a top-level agent often has several responsibilities. It may include user interface elements that are common to the entire application. In some systems, the presentation component does not exist.

The control component of the top PAC agent has three responsibilities:
- It allows the lower level agents to use the services of the top-level agent, mostly for access and manipulation of global data model. Input service requirements from lower level agents are passed to any Abstraction component or Presentation component.
- It coordinates the hierarchy of PAC agents. It maintains information on the relationships between top agents and lower-level agents. The control component uses this information to provide correct cooperation and exchange of data between the top-level agents and lower-level agents.
- It maintains information on user interaction with the system. For example, it can check whether a particular operation can be performed with the data model when started by the user. It can also keep track of functions that are called for provision of history and undo/redo service for operations with the functional core.

The low-level PAC agents are specific semantic concepts of the application domain, which can be as low as a simple graphic object, for example a circle, or as complex as a chart with strips that give a summary view of all data in the system. The Presentation component of the lowest PAC agent is a special view of the corresponding semantic concept and it provides access to all functions that users can apply. Internally, the Presentation component also maintains information about the view, such as position on the screen.

The Abstraction Component of the PAC agent lowest level has a similar responsibility as the Abstraction component at the PAC agent highest level, maintaining information that is specific to the agent. Contrary to the top agent abstraction component, other PAC agents do not depend on such data.

The control component of low-level PAC agent maintains the consistency between abstraction and presentation components, thereby avoiding a direct dependency between them. It serves as the adapter and executes the adaptation both of interface and data. This component of the lowest level PAC agents communicates with agents of higher level in order to exchange events and data. Incoming events, such as "close window" request - is sent to the presentation component of the low-level agent, while the incoming data are sent to its abstraction component. Outgoing events and information such as error messages are sent to an associated higher-level agent [8].

The top-level PAC agents are not limited only to providing semantic concepts of the application domain. The top-level agents can also be specified to implement system services. For example, the top-level agent may be the communication agent that allows the system to cooperate with other applications and to supervise the cooperation.

The middle-level PAC agents can fulfil two different roles: the composition and coordination. When, for example, each object in a complex graphic application is presented by a separate PAC agent, the middle-level agent groups such objects to form composite graphic object. Another role of the middle-level agent is maintaining consistency between the lower-level agents, for example, when carrying out coordination of multiple views on the same data.

PAC agent interfaces are designed using the "Compose Message" patterns. All input service requests, events and data are processed by one function called receiveMsg (). It interprets the messages and routes them to their recipient, which can be the agent abstraction or presentation component or another agent. Similarly, the function sendMsg () is used for packaging and delivery of events, data and services requests to other agents.

The PAC design pattern can be used in the design of context-aware mobile applications in the domain of PIM (Personal Information Management), which in addition to user input that comes through the phone's keyboard, need to accept incoming phone call as an input event. Via the controller, each input event causes the performance of actions on the model and consequent view update. J2ME platform does not offer APIs to control the voice communication and if we want to access these phone functions in terms of programming, it is necessary to use native applications. Therefore, it is necessary to achieve communication between the code for the Java 2 Mobile Edition and native programming code.

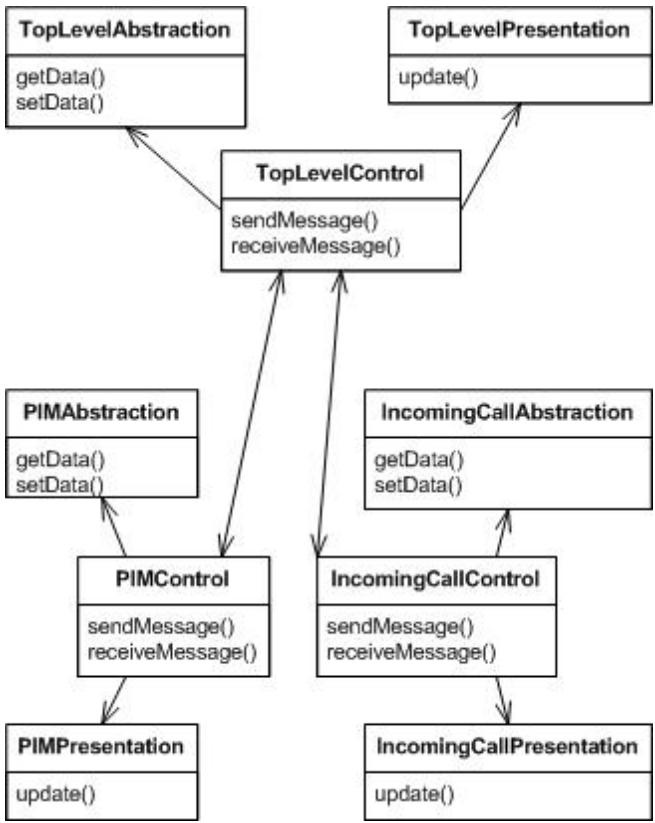

Fig. 3. Sequence Diagram as per PAC pattern

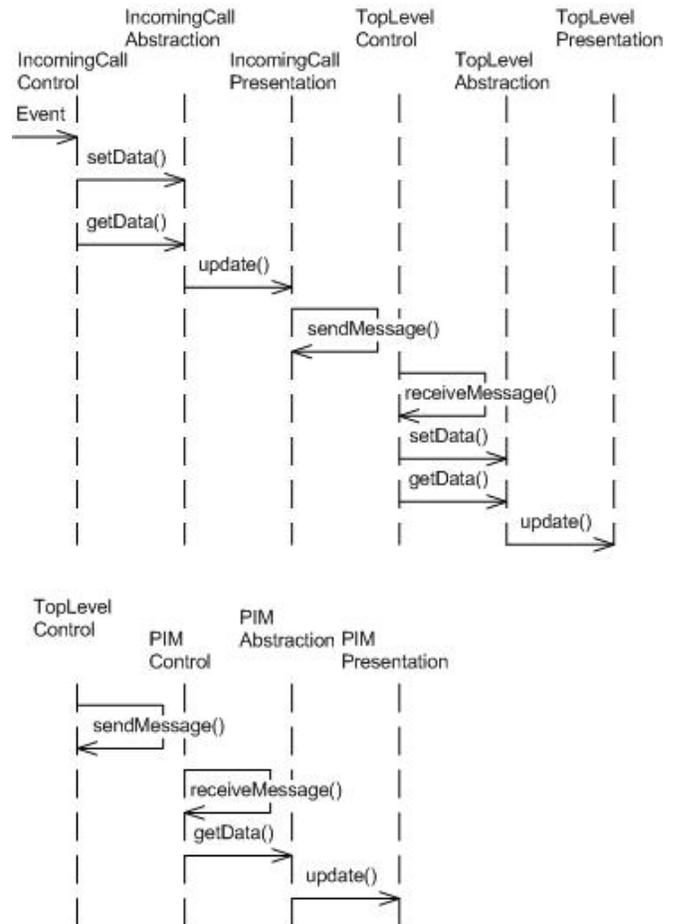

Fig. 4. Class diagram designed as per PAC pattern

On the Symbian operating system for mobile phones we can use sockets as a communication mechanism. By using the PAC design pattern, this communication mechanism will be used for communication between agents, namely their Control components. User application interface consists of the top agent whose function is to coordinate subordinate agents and the access to common data model and system behaviour. One of the subordinate agents accepts user input from the keyboard and another agent accepts incoming calls. Each of the subordinate agents participates in the updating of views. Given the structure of the PAC agent, communication controllers can act as clients and as servers. Therefore, the controller must have two threads, one to accept incoming communications from other controllers and one for initiating communication with other controllers. The Class diagram in Figure 3 and the Sequence diagram in Figure 4 show the proposed solution.

### 4.2 PAC architectural patterns benefits

The first benefit is the separation of tasks. The different semantic concepts in the application domain are presented by separate agents. Each agent maintains its own state and data, coordinated with, but independently of the other PAC agents. Individual PAC agents also offer their own human-computer interaction. This allows the development of dedicated data model for each semantic concept or task within the application, independently of other semantic concepts or tasks. Second, the PAC pattern provides support for change and expansion. Changes in the presentation or abstraction components of the PAC agent do not affect other agents in the system. This allows for modification of the data model in the base of the PAC agent or modification of its user interface, for example, from the command shell in the menus and dialogues. Furthermore, new agents easily integrate into existing PAC architecture without major changes of the existing PAC agents. All PAC agents communicate with each other through predefined interfaces. Additionally, existing agents can dynamically register the new PAC agents to provide communication and cooperation. Functionality to manage the new PAC agent such as a view coordination mechanism and changes and events propagation mechanism already exist.

The mode of coordination of agents in the PAC pattern provides a good basis for multitasking. PAC agents can be easily distributed to different threads, processes or machines. PAC agent extension with the appropriate functionality for inter-process cooperation only affects its control component.

## 4.2  PAC pattern disadvantages

The disadvantages of the PAC pattern are reflected in the increased complexity of the system. Implementation of each semantic concept as its own PAC agent can result in a complex system structure. If any graphic object, such as circle and rectangle, is implemented as a separate PAC agent, the system will drown in a sea of agents. Agents also must be coordinated and controlled, which requires additional coordinating agents. The level of design granulation should be carefully considered and also the point where to stop with dissolution of agents in an increasing number of lowest level agents.

In the PAC system, the control components are communication mediators between abstraction and presentation agent components and among different PAC agents. The quality of implementation of the control component is therefore critical both for effective cooperation among agents and the entire system architecture quality. Individual roles of the control component should be strictly separated from each other. Implementation of these roles should not be dependent on the specific details of other agents, such as their specific names or physical locations in the distributed system. Control component interface should be independent of internal details in order to secure for the agent's associates not to depend on the specific interfaces of their presentation or abstraction components. Responsibility of control component is to perform any adaptation of interfaces or data.

Additional tasks related to communication between PAC agents can affect the efficiency of the system. For example, if the lowest agent reads data from the top agent, all agents who are on the path from the lowest to the highest in the PAC hierarchy are included in this data exchange. If the agents are distributed, the data transfer also requires inter-process cooperation, which includes a series of actions that further burden the communication system.

## 5  TECHNIQUES TO MEET THE MOBILITY-ARISING DEMANDS

The main problem that arises with the growing number of platforms and interaction modalities is the need to create more application presentation components. One approach to avoid an increase in presentation components that must be developed by the programmers is the presentation transcoding [3]. The basic idea in transcoding is re-use of the View component for another device by using the automated transcoder that makes its transforming during execution. The transcoding technique focuses on the separation of information from the view in order to create indirect format that can be used for the production of other views. The process of creating this interim format is called transcoding. However, transcoding is not widely adopted because it fundamentally does not control the deeper structures and application semantics.

Transformation is another technique for limiting the growth of the view. It consists in describing the intent behind the user's interaction within the view component, instead of giving a real physical representation of the user interface management [3]. For example, the fact that the application requires from users to enter their age is presented by general INPUT element with a general limit on the data interval. Then, based on the characteristics of the end device, usability considerations or user preferences, the adaptation machine determines whether the INPUT element should be implemented as a text field, electoral list or even as a voice input. Several device-independent presentations have been developed over the years, including the User Interface Markup Language (UIML), Abstract User Interface Markup Language (AUIML), XForms and Microsoft ASP.NET Mobile Controls [9]. Regardless of adaptation technique used, systems that support the device-independent views also provide a number of integrated development environments for writing device-independent content.

Both above named techniques can be used regardless if the application is designed with the MVC or PAC pattern. It should be noted that transcoding and transformation system output only solve the problem of publication of multiple types of user interfaces to the user. It also ignores the fact that input from the user can come from a variety of heterogeneous channels such as HTTP, WAP, VoIP or GSM.

Limiting the increase in the number of controllers is handled in such a way that a component controller in mobile applications designed by MVC pattern is implemented as an Application Controller. It contains the control flow, including data validation and error management, usually in the procedure of handling events. There are several reasons why the application controller is required in cases of multiple destination devices:

- Different devices have different input hardware, from the keyboard, indication device, the microphone on the PC, to a pair of buttons and rollers for scrolling on the watch.

- The application flow may be different on different devices. For example, an application that contains the safe transactions may not complete a transaction on the device that does not have corresponding level of infrastructure. Similarly, an application that supports rich content may elect to skip the sides of the devices that are not able to present rich content
- During the restoration and opening of the page that is independent of the device, the page can be divided into multiple pages dependent on the device that is too small to contain the entire page.

As a result, a complete solution for multiple destination devices must include the introduction of application controllers. Application management and maintenance become more difficult with this increase in the number of views and controllers, which, according to [10] is especially true with the application of MVC pattern, due to close connection between views and controllers and to the asymmetric nature of this pattern. In contrast to MVC pattern, the components of the PAC pattern are very disconnected and therefore PAC scales very well. Different components can be linked together in a very loose way, because it is quite legal for the controllers to communicate and cooperate with each other. Consequently, making of complicated user interfaces based on simple components is much more natural for the PAC than for the MVC, because the composition of agents can be performed without violating encapsulation. In this pattern the different parts can be completely separate processes. These properties are what makes the PAC more suitable for development of mobile applications with multiple input and output modalities [10]. PAC pattern provides a well-defined place to add a variety of infrastructure components that take account of the dimensions of mobility, which is the control component. This means that the control component can communicate with location sensitivity system, the voice recognition engine, voice synthesis engine and all other subsystems that should be used for control and production of our user interface without compromising the separation of tasks between the abstraction and presentation and thereby protecting the business logic.

Given the conditions that led to the formation of PAC patterns, it can be concluded that the requirements of mobile applications just create such conditions that are best solved by using the PAC pattern. Variety of user interfaces, the existence of multiple interfaces, as well as the need for relative ease of adding new user interface to the system in the future, are all recognized as the requirements of mobile applications and the conditions that led to the recognition of PAC patterns as a solution. The first focus with this pattern is to ensure a uniform mechanism for communication between classes, as well as mechanisms for coordination structuring. This enables manageability of large number of components. The MVC pattern should be used in case that the application does not have a great number of types of user interfaces with different input-output modalities. In choosing an appropriate pattern for applications design with multiple types of user interfaces, we are guided by the avoidance of duplication of parts of the code and reduction of the number of classes and relationships between them to as a small number as possible.

Table 1: Comparison of MVC and PAC patterns

| Aspects | MVC versus PAC |
|---|---|
| Complexity | Greater complexity of PAC pattern because of multiple components and because of the way of implementation which may also include inter-process cooperation |
| Separation of tasks among the components of representation, control and pattern | MVC model can include interweaving of tasks, while PAC more strictly separates the tasks among the components |
| Efficiency | It may be lower with PAC pattern due to greater number of components and connections |
| Support to changes and extension | Due to interwoven tasks with MVC pattern, the advantage is on the side of the PAC pattern with a uniform structure |
| The existence of more substructures in the pattern, which require specific ways of interaction | More suitable PAC pattern |
| Control component complexity | Greater complexity of PAC pattern, due to need for communication with other control components |

In order to have correct interpretation of the comparisons given in Table 1, it should be noted that the need for communication with other control components, significantly makes complex the use of PAC design pattern. Because of this complexity, it is rarely used in mobile applications.

Although the MVC design pattern is widely used during the design of user interfaces in mobile applications, it is not universal. In certain circumstances, it is necessary to use a PAC design pattern, as it is primarily the case of requests for the existence of heterogeneous multimodal user interface. Such requests require the use of different technologies for the implementation of certain parts of the application, which can integrate with each other only by inter-process communication. In such cases it is impossible to use the MVC design pattern.

In [2] it is said that the PAC design pattern has the advantage that allows multitasking. However, the J2ME platform allows an application with multiple threads, so that this platform enables the controller to accept user input during its performance, for example, the HTTP communication. Therefore, the PAC has no exclusivity over multitasking.

# 6   CONCLUSION

The need for the existence of multiple input and output modalities in mobile applications leads to more complex user interfaces of mobile applications. Therefore, we should consider the available software patterns for building user interfaces in order to use more efficient structure for given requirements.

The paper reviews MVC and PAC software patterns and makes connection between the requirements of mobile application and features of such software patterns. The advantages and disadvantages of these software patterns are considered during their use in certain circumstances. This facilitates the choice of the software designers of the appropriate software pattern for building mobile application user interfaces.

**Plakalović D.** is a MSc student at the Faculty of Organizational Sciences, University of Belgrade. He is working at Faculty of Economics, University of Eastern Sarajevo, as assistant of professor. His interests are software development and applied information technologies.
**Simić D**, PhD, is a professor at the Faculty of Organizational Sciences, University of Belgrade. He received the B.S. in electrical engineering and the M.S. and the Ph.D. degrees in Computer Science from the University of Belgrade. His main research interests include: security of computer systems, organization and architecture of computer systems and applied information technologies.